\documentclass[pra,aps,eqsecnum,showpacs]{revtex4}
\usepackage{amssymb}

\usepackage{dcolumn}
\usepackage{amsmath}
\usepackage{graphics}

\usepackage[cp1251]{inputenc}
\usepackage[english]{babel}
\usepackage{amssymb,amsmath,epsfig}
\usepackage{graphicx}
\usepackage{mathtext}

\begin{document}
\renewcommand{\theequation}{\thesection.%
\arabic{equation}}
\title{Microscopic Theory of Scattering of Weak Electromagnetic Radiation by a Dense Ensemble of Ultracold Atoms}
\author{I.M.Sokolov${}^{1,2}$, D. V. Kupriyanov${}^{1}$}
\affiliation{$^{1}$St. Petersburg State Polytechnic University, St. Petersburg, 195251 Russia  \\ $^{2}$Institute for Analytical Instrumentation RAS,
 St. Petersburg, 198103 Russia } \email{IMS@IS12093.spb.edu }
\author{M.D. Havey}
\affiliation{Department of Physics, Old Dominion University, Norfolk, VA 23529}

\begin{abstract}
Based on the developed quantum microscopic theory, the interaction of weak electromagnetic radiation with dense ultracold atomic clouds is described in detail. The differential and total cooperative scattering cross sections are calculated for monochromatic radiation as particular examples of application of the general theory. The angular, spectral, and polarization properties of scattered light are determined. The dependence of these quantities on the sample size and concentration of atoms is studied and the influence of collective effects is analyzed. 

\end{abstract}

\maketitle

\section{Introduction}

A great number of experimental and theoretical papers devoted to the study of cold and ultracold atomic ensembles are related to the unique physical properties inherent in these systems and their broad possible practical applications. Improvements in techniques for cooling atomic gases in atomic traps \cite{1,2}  make their use very promising in metrology, frequency standards, and quantum information problems  \cite{3}-\cite{11}.  Most of the methods for diagnostics of the states of these ensembles are based on their interaction with electromagnetic radiation.  More importantly, the largest number of the many applications envisioned for ultracold atom ensembles have at their foundation such interactions.  

The interaction of radiation with cold atomic clouds reveals a number of specific features. First, atoms cooled down to low temperatures have very high resonance scattering cross sections, and therefore multiple scattering should be taken into account even in quite rarefied clouds. Second, atomic clouds in traps have nonequilibrium distributions over different hyperfine sublevels and inhomogeneous spatial distributions. Third, because of low mean velocities of atoms, interference effects can be observed during multiple scattering despite the absence of any ordering in the atomic cloud.

Depending on the concentration $n$ of atoms in a cloud, the cloud can interact with light either in the weak or strong localization regimes. The weak localization regime assumes that the medium is rarefied so that both the mean interatomic distances and the photon mean free path greatly exceed the light wavelength $\lambda $ In this case, individual atoms scatter radiation independently and multiple scattering can be treated as a number of successive scatterings by free atoms. By now, this regime has been studied in detail, both experimentally and theoretically \cite{20}-\cite{23}.

The interaction of light with dense atomic clouds has been studied far less thoroughly than with rarefied clouds. This is mainly caused by considerable technical difficulties associated with producing such ensembles under laboratory conditions. Some problems also arise in the consistent
quantum mechanical description of their interaction with the electromagnetic field. Under conditions when the mean distance between atoms becomes comparable with the light wavelength, it is necessary to consider the resonance interatomic dipole-dipole interaction. Because of this interaction, individual atoms cannot be treated as independent sources of secondary waves. Conditions are produced for the so-called dependent or cooperative scattering, and we are in fact dealing with the interaction of light with a giant quasimolecule combining all atoms in the ensemble.
Despite these difficulties, the strong localization regime has been attracting considerable attention recently because there is good reason to believe that interesting quantum processes are possible in this regime, such as the Anderson (strong) localization of light \cite{24}-\cite{27} and lasing in a disordered medium \cite{28}-\cite{32}, which can have practical applications. The aim of recent studies in this field was to obtain these effects experimentally and describe them theoretically.

The present paper is devoted to the theoretical study of light scattering by dense atomic clouds under conditions of strong localization, i.e., to developing a theory of dependent scattering. Obviously the problem of a dense ensemble belongs to the field of macroscopic electrodynamics. It is known that such problems can be solved by using either macroscopic or microscopic approaches. In the former case, description is performed by using such averaged characteristics as the field strength and atomic polarization, while the microscopic approach is based on the concept of the discrete structure of matter consisting of individual atoms or molecules. The macroscopic approach, which we employed to solve the scattering problem, was described in detail in \cite{SKKH09}. In the present paper, we develop the consistent quantum-mechanical description of the strong localization regime. We consider the scattering of quasiresonance radiation by an ultracold disordered ensemble of atoms under conditions when the mean number of atoms $\lambdabar^{3}\equiv k^{-3}$ in the volume can become comparable with unity. We assume that probe radiation is weak and restrict ourselves to the case when the field subsystem contains no more than one photon. Atoms are assumed immobile.

A key problem in describing scattering in this case is analysis of the evolution of an atomic ensemble after its excitation by initial radiation and investigation of possible decay channels for this radiation accompanied by the creation of secondary scattered photons. In this connection, the given problem is closely related to that of collective spontaneous decay in a dense polyatomic system. Moreover, this problem can be reduced to that of spontaneous decay by considering the evolution of an atomic subsystem in which only one atom located far from the cloud is excited in the initial state. By calculating the dynamics of such a system, we automatically describe how a photon emitted by this individual atom is scattered in the cloud.

The consideration of the problem in this formulation allows us to use methods developed for solving problems of superradiance of a singly excited ensemble. How dipole-dipole interaction influences collective spontaneous decay in a finite system was first studied for the case of two atoms in \cite{Stephan64} and  \cite{HH64}. A consistent quantum-mechanical approach developed in these papers was later generalized in a number of papers to polyatomic ensembles (see \cite{Lehmberg70}-\cite{Sw} and references therein). However, almost all papers on polyatomic systems used only a two-level atom model. This prevents the correct consideration of the polarization properties of light and, as a consequence, adequate description of the resonance dipole-dipole interaction at small distances. In addition, a real disordered atomic ensemble was replaced in calculations, as a rule, by a model of an averaged continuous medium. In this case, the role of random inhomogeneities was neglected. In our opinion, these simplifications considerably restrict the application of these theories for describing experiments with ultracold gases; therefore, we do not use them in our paper. Another important difference of our approach is that it can be used to analyze the scattering of coherent pulsed radiation, which is also important for interpretation of experiments.

Our paper contains four sections, conclusions, and an appendix. In Section 2, we solve the nonstationary Schrodinger equation to calculate the amplitudes of a number of states of a combined system including atoms and electromagnetic radiation. In Section 3, the calculated amplitudes are used to determine the intensity, spectrum, and polarization properties of scattered radiation. Section 4 contains the description and results of numerical calculations. In the Conclusions, the main results are briefly formulated and the possibilities for further investigations in this field are discussed.

\section{STATE AMPLITUDES}
\setcounter{equation}{0}

The state of a closed system consisting of atoms and electromagnetic field can be described by the wavefunction $\psi$. This function in the Schrodinger representation satisfies the equation
\begin{equation} i\hbar \frac{\partial \psi
}{\partial t}=H\psi ,  \label{1}
\end{equation} 

\begin{equation}
H=H_{0}+V=H_{f}+\sum\limits_{a}H_{a}+V.  \label{2}
\end{equation}

Here $H_{a}$, $H_{f}$ and $V$ are Hamiltonians of free atoms not interacting with each other and the free field, and their interaction operator, respectively. In the dipole approximation used here, we have
\begin{equation}
V=-\sum_{a}\mathbf{d}^{(a)}\mathbf{E}(\mathbf{r}_{a}),  \label{3}
\end{equation}
where $\mathbf{E}(\mathbf{r})$ -- is the electric displacement vector operator \footnote{In the literature, this operator is often called the field strength operator or simply the field operator. Note, however, that in the dipole approximation used here it coincides with the field strength operator only outside the atomic medium under study. This difference is discussed in detail, for example, in  \cite{ChT}.}.

\begin{equation}
\mathbf{E}(\mathbf{r})=\mathbf{E}^{(+)}(\mathbf{r})+\mathbf{E}^{(-)}(\mathbf{r})=
i\sum\limits_{\mathbf{k},\alpha }\sqrt{\frac{2\pi \hbar \omega
_{k}}{V}}\mathbf{e}_{\mathbf{k}\alpha }a_{\mathbf{k}\alpha }\exp (i\mathbf{kr})+ h.c.  \label{4}
\end{equation}
$\mathbf{E}^{(\pm )}-$ are the operators of positive and negative frequency components;
$a_{\mathbf{k}\alpha }$ -- is the photon annihilation operator in a mode with wave vector $\mathbf{k}$ and polarization  $\alpha ;$ $V$ is the quantization volume; is the dipole moment operator of the atom $a$, $\mathbf{e}_{\mathbf{k}\alpha }$ are polarization unit vectors satisfying the transversability condition;
\begin{equation}
\sum\limits_{\alpha }\mathbf{e}_{\mathbf{k}\alpha }^{\mu }\mathbf{e}_{\mathbf{k}\alpha }^{\ast \nu }=
\delta _{\mu \nu }-\frac{k_{\mu }k_{\nu }}{k^{2}}.  \label{5}
\end{equation}

We will solve equation (\ref{1}) by using the general approach developed in Heitler's papers and considered in detail in \cite{G}. We will seek the wavefunction $\psi $ in the form of the expansion in a set of eigenstates $\left\{ |l\rangle \right\} $ of the operator $H_{0}$:
\begin{equation} \psi =\sum_{l}b_{l}(t)|l\rangle .
\label{6}
\end{equation}
Here, the subscript  $l$  defines the state of all atoms and the field.

Equation (\ref{1}) gives the system of coupled equations
\begin{equation} i\hbar \frac{\partial b_{l}(t)}{\partial
t}-E_{l}b_{l}(t)=\sum\limits_{j}V_{lj}b_{j}(t).  \label{7}
\end{equation}
for the state amplitudes $b_{j}(t)$. Here $E_{l}$ is the energy of the lowest state of the system of free atoms and field.

System (\ref{7}) contains an infinite number of equations and cannot be solved in the general case. In this paper, as mentioned above, we will study the evolution of only weakly excited states of the combined system. In addition, we will calculate all the observables of interest, namely, scattering cross sections, decay constants, levels shifts, etc., correct to the second order in the fine structure constant. The higher-order radiative corrections will not be calculated. In this case, the number of $|l\rangle$ states taken into account in calculations can be restricted. As shown in \cite{Stephan64}, in this approximation it is sufficient to consider only the states
\begin{eqnarray} \psi _{g} &=&|g,g,...g\rangle \otimes
|\mathbf{k}\alpha \rangle
;\;E_{g}=\hbar \omega _{k};  \notag \\
\psi _{e_{a}} &=&|g,g,...g,e,g,...g\rangle \otimes |vac\rangle
;\;E_{e_{a}}=\hbar \omega _{a}.  \label{8} \\
\psi _{e_{a}e_{b}} &=&|g,...g,e,g...g,e,g,...g\rangle \otimes |\mathbf{k}\alpha \rangle ;\;
E_{e_{a}e_{b}}=\hbar (\omega _{a}+\omega _{b})+\hbar
\omega _{k}.  \notag
\end{eqnarray}
Here, $\omega _{a}$ is the transition frequency of the atom $a$ and the ground state energy of the atoms is set equal to zero.

Note that, by choosing the states considered in the calculations, we go beyond the rotating wave approximation because we included states with energy $\hbar (\omega _{a}+\omega_{b})+\hbar \omega _{k},$ which contain, along with a photon, two excited atoms $a$ and $b.$ The consideration of nonresonance states proves necessary to describe correctly the dipole-dipole interaction of atoms separated by small distances less than the light wavelength. Indeed, the dipole-dipole interaction can be described as the result of exchange by resonance photons. Such an exchange can occur in two ways, either via resonant or nonresonant intermediate states. In the first case, an excited atom emits a photon, which then is absorbed by an unexcited atom. In the second case, at first, an atom is excited and emits a photon and then the initially excited atom undergoes a transition to the ground state and absorbs a photon.
According to the energy-time uncertainty relation, during the propagation time of $\tau =r/c$ a photon between two atoms, virtual states can exist with an energy that differs from the resonance (initial) energy by the value $\Delta E\sim \hbar c/r.$  When the interatomic distance r is large, it turns out that the second way is strongly nonresonant and therefore is usually neglected. If the interatomic distance is smaller or close to the reciprocal wave vector $r\lesssim \lambdabar $ ,then $\ \tau \lesssim \lambdabar /c=1/\omega $ and $\Delta E\gtrsim \hbar ck =\hbar \omega$. Therefore, we should take into account the second energy exchange channel as well, i.e. the channel involving the intermediate state $\psi _{e_{a}e_{b}}.$. The applicability of the rotating wave approximation in problems of the spontaneous decay of excited states of polyatomic systems is considered in more detail, for example, in \cite{FrM} -\cite{Sw}.

Below, we will consider, along with the spontaneous decay and scattering of one-photon radiation, the interaction of an atomic ensemble with weak coherent radiation. It is known that weak coherent radiation can be quite accurately represented as the superposition of one-photon and vacuum states. Therefore, we will add the state
\begin{equation} \psi _{g^{\prime }}=|g,g,...g,g\rangle
\otimes |vac\rangle ;\;E_{g^{\prime }}=0.  \label{8.2}
\end{equation}
to group (\ref{8}).
Further simplifications will be performed for a system consisting of $N+1$ atoms. We assume that $N$ atoms forming a cloud are identical and can be described by a four-level system with one ground and three degenerate excited Zeeman sublevels (the $J=0\leftrightarrow 
J=1$ transition). One of the four-level atoms, which we will call a source, can differ in the transition frequency and the excited-state lifetime. The general expressions obtained below will be also valid in a particular case when all $N+1$ atoms are identical and form one cloud.

Consider the dynamics of a combined atom + field system, assuming that all atoms at the initial moment, except the source, are in the ground state, while the field is in the vacuum state. The source atom is in the superposition of the ground and excited states. This means that at the initial moment two states are excited, one of which is $g^{\prime }$, and the second state is denoted by $o$. If $b_{o}$ is the initial amplitude of the $o$ state and $b_{o^{\prime }}$ is that

of the $g^{\prime }$ state $(\left\vert b_{o}\right\vert ^{2}+\left\vert
b_{o^{\prime }}\right\vert ^{2}=1)$, then system of equations (\ref{7}) for amplitudes should be supplemented by the initial conditions
\begin{equation}
b_{i}(0)=0;\;b_{o}(+0)=b_{o};\;b_{g^{\prime }}(+0)=b_{o^{\prime
}};\;i\neq o,g^{\prime }.
\end{equation}
The solution obtained in this case describes the state of the atomic-field system for positive times $t>0$.

Obviously the amplitude of the state  $\psi _{g^{\prime }}$ (\ref{8.2}) in the approximations used does not change during the evolution of the system. This is explained by the fact that transitions to this state from other states taken into account in the calculations are impossible. The transition from $g^{\prime }$ to any of the states (\ref{8}) is also impossible (virtual transitions to a state with one photon and one excited atom, which cause Lamb shifts, are taken into account by the renormalization of the transition frequency; see the Appendix), and therefore $b_{g^{\prime }}(t)=b_{o^{\prime }}$.

To determine the time dependence of other amplitudes, we extend, by the Heitler method, the solution of system (\ref{7}) to the negative time semiaxis, assuming that

\begin{equation*} b_{i}(t)=0;\;t<0.
\end{equation*}
The function $b_{o}(t)$ has a jump for $t=0.$ We will take this into account by adding the inhomogeneous term into the equation for $b_{o}(t)$, 
\begin{equation}
i\hbar \frac{\partial b_{l}(t)}{\partial t}-E_{l}b_{l}(t)=\sum\limits_{j}V_{lj}b_{j}(t)+i\hbar
\delta _{lo}\delta (t)b_{o}.  \label{9}
\end{equation}
where $\delta_{ik}$ is the Kronecker delta. The system obtained in this way is solved by
passing to the Fourier transform

\begin{equation*}
b_{l}(t)=\int\limits_{-\infty }^{\infty }\dfrac{d\omega }{2\pi }b_{l}(\omega
)\exp (-i\omega t).
\end{equation*}
In this case, 
\begin{equation}
\left( \hbar \omega -E_{l}\right) b_{l}(\omega
)=\sum\limits_{j}V_{lj}b_{j}(\omega )+i\hbar \delta _{lo}b_{o}.
\label{11}
\end{equation}
We will separate from (\ref{11}) the equation for $b_o$, taking into account that $V_{oo}=0.$
\begin{eqnarray}
\left( \hbar \omega -E_{o}\right) b_{o}(\omega )
&=&\sum\limits_{j\neq
g}V_{oj}b_{j}(\omega )+i\hbar b_{o}.  \label{12.1} \\
\left( \hbar \omega -E_{l}\right) b_{l}(\omega )
&=&V_{lo}b_{o}(\omega )+\sum\limits_{j\neq o}V_{lj}b_{j}(\omega ).
\label{12.2}
\end{eqnarray}
We introduce new variables  $u_{l}(\omega )$
\begin{equation}
b_{l}(\omega )=u_{l}(\omega )b_{o}(\omega )\varsigma \left( \hbar
\omega -E_{l}\right) .  \label{13}
\end{equation}
By  introducing a singular function $\varsigma \left( x\right)
=\underset{k\rightarrow \infty }{\lim }(1-\exp (ikx))/x$ , it is possible to prove in the general case \cite{G} that the obtained solution will satisfy the initial conditions. 

Functions $u_{l}(\omega )$ satisfy the system of equations
\begin{eqnarray}
u_{g}(\omega ) &=&V_{go}+\sum\limits_{e\neq o}V_{ge}u_{e}(\omega )\varsigma
\left( \hbar \omega -E_{e}\right) ,  \notag \\
u_{ee}(\omega ) &=&V_{ee;o}+\sum\limits_{e\neq o}V_{ee;e}u_{e}(\omega
)\varsigma \left( \hbar \omega -E_{e}\right) ,  \label{15} \\
u_{e}(\omega ) &=&\sum\limits_{g}V_{eg}u_{g}(\omega )\varsigma \left( \hbar
\omega -E_{g}\right) +\sum\limits_{ee}V_{e;ee}u_{ee}(\omega )\varsigma
\left( \hbar \omega -E_{ee}\right) ,\;e\neq o,  \notag
\end{eqnarray}
By solving this system, we obtain the amplitude $b_{o}(\omega )$
\begin{gather}
b_{o}(\omega )=\frac{ib_{o}}{\omega -\omega _{o}-\Sigma _{o}(\omega )},
\label{14.2} \\
\Sigma _{o}(\omega )=\Delta -i\Gamma /2=\sum\limits_{g}V_{og}u_{g}(\omega
)\varsigma \left( \hbar \omega -E_{g}\right) /\hbar
+\sum\limits_{ee}V_{o;ee}u_{ee}(\omega )\varsigma \left( \hbar \omega
-E_{ee}\right) /\hbar  \label{14.3}
\end{gather}
The subscript of the excited atom is omitted for brevity in relations (\ref{15})-(\ref{14.3}), and it is also taken into account that the relation $x\varsigma \left( x\right) =1$ is fulfilled for singular functions $\varsigma \left( x\right) $.

By using the first and second equations of system (\ref{15}), we obtain the closed system of equations
\begin{gather} u_{e}(\omega
)=\sum\limits_{g}V_{eg}V_{go}\varsigma \left( \hbar \omega
-E_{g}\right) +\sum\limits_{ee}V_{e;ee}V_{ee;o}\varsigma \left(
\hbar \omega
-E_{ee}\right) +  \notag \\
+\sum\limits_{e^{\prime }\neq o}\left[ \sum\limits_{g}V_{eg}V_{ge^{\prime
}}\varsigma \left( \hbar \omega -E_{g}\right)
+\sum\limits_{ee}V_{e;ee}V_{ee;e^{\prime }}\varsigma \left( \hbar \omega
-E_{ee}\right) \right] u_{e^{\prime }}(\omega )\varsigma \left( \hbar \omega
-E_{e^{\prime }}\right) ,\;e\neq o.  \label{16}
\end{gather}
for the states with one excited atom in the cloud. 

Note that this system, unlike (\ref{9}), contains a finite number of equations. This number is determined by the number of atoms in the cloud and the number of excited states considered in the calculations. The sums $\sum\limits_{g}V_{eg}V_{ge^{\prime }}\varsigma \left( \hbar \omega
-E_{g}\right) $ and $\sum\limits_{ee}V_{e;ee}V_{ee;e^{\prime }}\varsigma \left( \hbar \omega -E_{ee}\right) $
 can be calculated for an infinitely broadband field reservoir by assuming that the quantization volume is infinite (see the Appendix). By using expressions presented in the Appendix, we rewrite system
(\ref{16}) in a more compact form:
\begin{equation}
\sum\limits_{e^{\prime }\neq o}\left[ (\omega -\omega _{a})\delta
_{ee^{\prime }}-\Sigma _{ee^{\prime }}(\omega )\right] u_{e^{\prime
}}(\omega )\varsigma \left( \hbar \omega -E_{e^{\prime }}\right) =\Sigma
_{eo}(\omega )  \label{19}
\end{equation}
Here, the vector $\Sigma _{eo}(\omega )$ in the right-hand side determines the excitation of different atoms in the cloud by the source radiation, while the matrix $\Sigma _{ee^{\prime }}(\omega )$ describes radiation transferred between atoms in the cloud.

System (\ref{19}) can be reduced to an integral equation by using the continuous medium approximation. This significantly simplifies the solution of the problem for a two-level atom system \cite{FrM}-\cite{Sw}. Moreover, in this case, even an analytic solution is possible for spatially homogeneous spherical clouds. This solution neglects, however, the important properties of real physical systems, and therefore we will solve the linear system (\ref{19}) numerically. The numerical solution is restricted by the possibilities of available computers which allow us to consider clouds containing a few thousands of atoms. The number of atoms in quasi static traps, where the required concentrations can be achieved, can be two orders of magnitude greater; however, by using this approach, we can correctly describe all polarization effects taking into account random inhomogeneities of the medium. Both these factors, as follows from our previous calculations \cite{SKKH09}, are quite important.

The formal solution of system (\ref{19}) can be written in the form
\begin{equation}
u_{e}(\omega )\varsigma \left( \hbar \omega -E_{e}\right) =\sum\limits_{e^{\prime }\neq
o}R_{ee^{\prime }}(\omega )\Sigma _{e^{\prime }o}(\omega ),  \label{21}
\end{equation}
where
\begin{equation}
R_{ee^{\prime }}(\omega )=\left[ (\omega -\omega _{a})\delta _{ee^{\prime }}-\Sigma
_{ee^{\prime }}(\omega )\right] ^{-1}.  \label{21.1}
\end{equation}

By using (\ref{21}) and (\ref{21.1}, we obtain the amplitudes of all states of interest for us: 
\begin{eqnarray}
b_{g^{\prime }}(t) &=&b_{o^{\prime }};  \label{22.3} \\
b_{o}(t) &=&\int\limits_{-\infty }^{\infty }\dfrac{id\omega }{2\pi }\frac{b_{o}\exp (-i\omega t)}{\omega -\omega _{o}-
\Sigma _{o}(\omega )};
\label{22.4} \\
b_{g}(t) &=&\int\limits_{-\infty }^{\infty }\dfrac{id\omega }{2\pi }
\frac{b_{o}\varsigma \left( \hbar \omega -E_{g}\right) \exp (-i\omega t)}{\omega
-\omega _{o}-\Sigma _{o}(\omega )}\left( V_{go}+\sum\limits_{e,e^{\prime
}\neq o}V_{ge}R_{ee^{\prime }}(\omega )\Sigma _{e^{\prime }o}(\omega
)\right) ;  \label{23} \\
b_{e}(t) &=&\int\limits_{-\infty }^{\infty }
\dfrac{id\omega }{2\pi }\frac{b_{o}\exp (-i\omega t)\sum\limits_{e^{\prime }\neq o}R_{ee^{\prime }}(\omega
)\Sigma _{e^{\prime }o}(\omega )}{\omega -\omega _{o}-\Sigma _{o}(\omega )};
\label{24} \\
b_{ee}(t) &=&\int\limits_{-\infty }^{\infty }\dfrac{id\omega }{2\pi }\frac{b_{o}\varsigma
\left( \hbar \omega -E_{ee}\right) \exp (-i\omega t)}{\omega
-\omega _{o}-\Sigma _{o}(\omega )}\left( V_{ee;o}+\sum\limits_{e,e^{\prime
}\neq o}V_{ee;e}R_{ee^{\prime }}(\omega )\Sigma _{e^{\prime }o}(\omega
)\right) ;  \label{25} \\
\Sigma _{o}(\omega ) &=&\Sigma _{oo}(\omega )+\sum\limits_{e,e^{\prime }\neq
o}\Sigma _{oe}(\omega )R_{ee^{\prime }}(\omega )\Sigma _{e^{\prime
}o}(\omega )  \label{26}
\end{eqnarray}

Relations (\ref{22.3})-(\ref{26}) completely describe the studied system under the above assumptions. In the next section, we will use them to obtain explicit expressions for scattered radiation parameters observed in experiments. However, an important preliminary remark should be made. Expressions (\ref{22.3})-(\ref{26}) are valid for any distance between an atom excited from the ground state and the cloud. They describe, in particular, the modification of the spontaneous decay of the atom near the cloud or even inside it. Our main goal is to describe the scattering of light by a dense atomic cloud. Therefore, below we restrict ourselves to the case when a source atom is located far from the cloud. In this case, we can neglect the reverse influence of the cloud on the spontaneous decay of this isolated atom taking into account only the first term in (\ref{26}) by calculating $\Sigma _{o}(\omega )$. By including the Lamb shift in the definition of frequency $\omega _{o}$, we obtain
\begin{equation} \Sigma _{o}(\omega )=-i\gamma _{o}/2.  \label{27}
\end{equation}
Here, $\gamma _{o}$ is a decay constant of the atom that was initially excited. This constant can be arbitrary.

Along with simplification of the quantity $\Sigma _{o}(\omega )$, the expression for the vector $\Sigma _{eo}(\omega )$ describing excitation of different atoms in the cloud by the source is also simplified in this case. Because the distance from the cloud to the emitting atom is large, retardation effects are important and the polar approximation cannot be used. However, we can use the rotating wave approximation and retain only one term decreasing the least rapidly with increasing distance from the source to the cloud. In this case
\begin{equation} \Sigma _{eo}(\omega
)=-\sum\limits_{\mu ,\nu }\frac{\mathbf{d}_{e;g}^{\mu
}\mathbf{d}_{g_{o};e_{o}}^{\nu }}{\hbar r_{eo}}\left[ \delta _{\mu
\nu }-\dfrac{\mathbf{r}_{eo_{\mu }}\mathbf{r}_{eo_{\nu
}}}{r_{eo}^{2}}\right] \left( \frac{\omega }{c}\right) ^{2}\exp
\left( i\frac{\omega r_{eo}}{c}\right) ;  \label{27.1}
\end{equation}
Here, $\mathbf{r}_{eo}$ is the vector drawn from the source o$o$ to the atom excited to the $e$ state (hereafter for brevity, we will call it the $e$ atom). We will transform expression (\ref{27.1}) taking into account that $r_{eo}$ greatly exceeds the size of the system. We introduce a coordinate system with a center inside the cloud. Let $\mathbf{r}_{e}$ an $\mathbf{r}_{o}$ be the radius vectors of the cloud atoms and the source, respectively. Then,
$r_{oe}=\sqrt{\left(
\mathbf{r}_{o}-\mathbf{r}_{e}\right) ^{2}}\simeq
r_{o}(1-\mathbf{r}_{o}\mathbf{r}_{e}\mathbf{/}r_{o}^{2})=
r_{o}-\mathbf{nr}_{e}\mathbf{;\;n=r}_{o}/r_{o}$ and
\begin{equation} \Sigma _{eo}(\omega )=-\sum\limits_{\mu ,\nu
}\frac{k^{2}\mathbf{d}_{e;g}^{\mu }\mathbf{d}_{g_{o};e_{o}}^{\nu
}}{\hbar r_{o}}\left[ \delta _{\mu \nu }-\frac{\mathbf{k}_{\mu
}\mathbf{k}_{\nu }}{k^{2}}\right] \exp \left(
ikr_{0}-i\mathbf{kr}_{e}\right) ;  \label{27.2}
\end{equation}
Here, $\mathbf{k=}\omega \mathbf{n/}c$. We can see from (\ref{27.2}) that the source radiation in the atomic ensemble region can be treated quite accurately as a plane wave.

\section{TRANSITION PROBABILITIES AND SCATTERING CROSS SECTION}

\setcounter{equation}{0}
Knowing the amplitudes of the states, we can calculate the probability of finding the system in a specified state at any instant of time. In experiments, as a rule, the parameters of electromagnetic radiation are measured at large distances from a scattering cloud. These parameters are determined by the quantity $b_{g}(t),$. The calculation of this quantity is the main purpose of our paper because different states $g$ correspond to different frequencies, polarizations, and wave vectors of the emitted photon.
Relation (\ref{23}) allows us to find $b_{g}(t)$ at an arbitrary instant of time. Below we consider the two most important particular cases allowing a significant simplification in the analytic form. The first case is the calculation of the probability of different field states after a long time interval when spontaneous decay has ended. In our case, these are times exceeding the excited-state decay times of the source atom and atomic ensemble. The second case is the calculation of the transition probability per unit time when the system state still only weakly differs from the initial state. These times are considerably shorter than the decay time (but, of course, they greatly exceed reciprocal transition frequencies).

We will solve this problem taking $\underset{t\rightarrow +\infty }{lim}\varsigma \left( \hbar \omega
-E_{g}\right) \exp (-i\omega t)=-2\pi i\delta (\hbar \omega -E_{g})\exp
(-iE_{g}t/\hbar ).$ 
\begin{equation}
b_{g}(t)=\dfrac{1}{\hbar }\frac{b_{o}\exp (-i\omega _{g}t)}{\omega
_{g}-\omega _{o}-\Sigma _{o}(\omega _{g})}\left(
V_{go}+\sum\limits_{e,e^{\prime }\neq o}V_{ge}R_{ee^{\prime }}(\omega
_{g})\Sigma _{e^{\prime }o}(\omega _{g})\right) ,\;t\rightarrow +\infty
\label{28}
\end{equation}
Here  $\omega _{g}=E_{g}/\hbar $ is the frequency of the emitted photon. 

The first term in expression (\ref{28}) describes radiation incident directly on a photodetector from the initially excited atom. The angular and spectral distributions of radiation from the cloud are determined by the square of the modulus of the second term in parentheses in (\ref{28}). We will assume here that the photodetector is equipped with a set of apertures eliminating the contribution of the initially excited atom to the photocurrent. Then, the probability of finding a photon in the final state in a given mode will be
\begin{equation} P_{g}=\dfrac{1}{\hbar ^{2}}\frac{\left\vert
b_{o}\sum\limits_{e,e^{\prime }\neq o}V_{ge}R_{ee^{\prime }}(\omega
_{g})\Sigma _{e^{\prime }o}(\omega _{g})\right\vert ^{2}}{\left(
\omega _{g}-\omega _{o}\right) ^{2}+\gamma _{o}^{2}/4}  \label{29}
\end{equation}
The dominator in this expression is related to the efficiency of excitation of the cloud by the source radiation. We calculated it using expression (\ref{27}). For small $\gamma _{0}$, only a narrow spectral group of the cloud states is efficiently excited. To determine the total excitation spectrum of the cloud in this case, it is necessary to repeat experiments many times by changing the source radiation frequency $\omega _{o}$. We can obtain the cloud radiation spectrum in one experiment by using a radiation source emitting a very short pulse, so that $\gamma _{0},$ is large and all the characteristic collective excitation frequencies lie within the pulse spectral width $\gamma _{0},$.

In another limiting case, when $\gamma _{0},$ is much smaller than the characteristic rates of transient processes in the cloud, a quasi-stationary process will take place during which the quasi-stationary distribution of excitation over the cloud will be established. This distribution will then change slowly (with the decay rate $\gamma _{0}^{-1}$ of the excited state of the source atom). In this case, we can calculate the transition probability per unit time $w=d\left\vert b_{g}(t)\right\vert ^{2}/dt$. To do this, it is necessary to perform the passage to the limit $\gamma _{o}\rightarrow 0$ before passing to large times:
\begin{equation} b_{g}(t)\underset{\gamma _{0}\rightarrow
0}{\rightarrow }\int\limits_{-\infty }^{\infty }\dfrac{ib_{o}d\omega
}{2\pi \hbar }\varsigma \left( \omega -\omega _{g}\right) \varsigma
\left( \omega -\omega _{o}\right) \exp (-i\omega t)\left(
V_{go}+\sum\limits_{e,e^{\prime }\neq o}V_{ge}R_{ee^{\prime
}}(\omega )\Sigma _{e^{\prime }o}(\omega )\right) \label{31}
\end{equation}
We will calculate $w$ using the expansion of the product $\varsigma \left( \omega -\omega _{g}\right) \varsigma \left( \omega
-\omega _{o}\right) $ into simple fractions: $\varsigma _{\gamma
_{0}}\left( \omega -\omega _{o}\right) \varsigma _{\sigma }\left(
\omega -\omega _{g}\right) =\left[ \varsigma _{\gamma _{0}}\left(
\omega -\omega _{o}\right) -\varsigma _{\sigma }\left( \omega
-\omega _{g}\right) \right] \varsigma _{\sigma -\gamma _{0}}\left(
\omega _{o}-\omega _{g}\right) .$ Here $\sigma $ and $\gamma _{0}$ are constants. By having them tend to zero, we obtain the corresponding generalized function (further we will assume for definiteness that $\sigma $ > $\gamma _{0}$, which does not affect the final result). Taking into
account that $\varsigma _{\sigma -\gamma _{0}}\left( \omega
_{o}-\omega _{g}\right) -\varsigma _{\sigma -\gamma _{0}}^{\ast
}\left( \omega _{o}-\omega _{g}\right) =-2\pi i\delta \left( \omega
_{o}-\omega _{g}\right) $, we obtain after simple transformations
\begin{equation} w=2\pi \left\vert b_{o}\right\vert
^{2}\frac{\left\vert V_{go}+\sum\limits_{e,e^{\prime }\neq
o}V_{ge}R_{ee^{\prime }}(\omega _{o})\Sigma _{e^{\prime }o}(\omega
_{o})\right\vert ^{2}}{\hbar ^{2}}\delta \left( \omega _{o}-\omega
_{g}\right) . \label{32}
\end{equation}
As earlier, the first term describes the contribution of radiation emitted directly by the initially excited atom. 

Expressions (\ref{29})-(\ref{32}) can be used to describe experiments in which the occupation of different modes of the field is detected, for example, when modes with a narrow group of wave vector directions are selected with the help of a convergent lens. To describe experiments without such a selection, for example, with point photodetectors, it is necessary to know the mean parameters of the field at a certain spatial point at a specified instant of time. These mean parameters can be found from the known wave function
\begin{equation} \psi =b_{o}(t)\psi _{o}+b_{o^{\prime }}\psi
_{g^{\prime }}+\sum_{g}b_{g}(t)\psi _{g}+\sum_{e\neq o}b_{e}(t)\psi _{e}+\sum_{ee}b_{ee}(t)\psi
_{ee}.  \label{33}
\end{equation}
For example, the mean projection $\alpha $ of the vector $\mathbf{E}^{(+)}$ at the point $\mathbf{r}$ at the instant $t$ is 

\begin{equation} \left\langle \psi \right\vert
\mathbf{e}_{\alpha }^{\ast }\mathbf{E}^{(+)}(\mathbf{r})\left\vert
\psi \right\rangle =\int\limits_{-\infty }^{\infty }\dfrac{id\omega
}{2\pi }\frac{\hbar b_{o}b_{o^{\prime }}^{\ast }\exp (-i\omega
t)}{\omega -\omega _{o}+i\gamma _{o}/2}\left( \widetilde{\Sigma
}_{\alpha o}(\omega )+\sum\limits_{e,e^{\prime }\neq
o}\widetilde{\Sigma }_{\alpha e}(\omega )R_{ee^{\prime }}(\omega
)\Sigma _{e^{\prime }o}(\omega )\right) .  \label{34}
\end{equation}
Here, the matrix $\widetilde{\Sigma }_{\alpha e}(\omega )$ describes the propagation of radiation from an atom excited in the state $e$ to the observation point, 

\begin{equation} \widetilde{\Sigma }_{\alpha
e}(\omega )=\sum_{g}\left\langle o^{\prime }\right\vert
\mathbf{e}_{\alpha }^{\ast }\mathbf{E}^{(+)}(\mathbf{r})\left\vert
g\right\rangle V_{ge}\varsigma \left( \hbar \omega -E_{g}\right)
/\hbar .  \label{35}
\end{equation}
The explicit expression for this matrix can be obtained as it was done in the Appendix upon calculation of infinite sums entering equation (\ref{16}). In the rotating wave approximation, we obtain
\begin{equation} \widetilde{\Sigma }_{\alpha e}(\omega
)=-\frac{1}{\hbar \left\vert \mathbf{r-r}_{e}\right\vert }\left[
\mathbf{e}_{\alpha }^{\ast
}\mathbf{d}_{g;e}-\dfrac{(\mathbf{e}_{\alpha }^{\ast }\left(
\mathbf{r-r}_{e}\right) )\left( \mathbf{d}_{g;e}\left(
\mathbf{r-r}_{e}\right) \right) }{\left\vert
\mathbf{r-r}_{e}\right\vert ^{2}}\right] \left( \frac{\omega
}{c}\right) ^{2}\exp \left( i\frac{\omega \left\vert
\mathbf{r-r}_{e}\right\vert }{c}\right) . \label{36}
\end{equation}
Here, $\mathbf{r}_{e}$ is the radius vector of the atom excited to state $e$  and $\mathbf{r}$ is the radius vector of the observation point in the chosen coordinate system.

The polarization components of the intensity correlation function can be calculated similarly:
\begin{equation}
\left\langle \psi \right\vert \mathbf{e}_{\alpha }\mathbf{E}^{(-)}(\mathbf{r})\mathbf{e}_{\alpha }^{\ast }
\mathbf{E}^{(+)}(\mathbf{r})\left\vert \psi
\right\rangle =\left\vert b_{o}\right\vert ^{2}\left\vert
\int\limits_{-\infty }^{\infty }\dfrac{id\omega }{2\pi }\frac{\hbar \exp
(-i\omega t)}{\omega -\omega _{o}+i\gamma _{o}/2}\left( \widetilde{\Sigma }_{\alpha o}(\omega )+
\sum\limits_{e,e^{\prime }\neq o}\widetilde{\Sigma }_{\alpha e}(\omega )R_{ee^{\prime }}(\omega )
\Sigma _{e^{\prime }o}(\omega
)\right) \right\vert ^{2}.  \label{37}
\end{equation}

Expressions (\ref{34}) and (\ref{37}) are considerably simplified in the quasistatic case. By retaining only the contribution of light scattered by the cloud and passing to the limit $\gamma _{o}\rightarrow 0$
\begin{eqnarray}
\left\langle \psi \right\vert \mathbf{e}_{\alpha }^{\ast }\mathbf{E}^{(+)}(\mathbf{r})
\left\vert \psi \right\rangle &=&\hbar b_{o}b_{o^{\prime }}^{\ast
}\exp (-i\omega _{o}t)\sum\limits_{e,e^{\prime }\neq o}\widetilde{\Sigma }_{\alpha e}(\omega _{o})
R_{ee^{\prime }}(\omega _{o})\Sigma _{e^{\prime
}o}(\omega _{o});  \label{38} \\
\left\langle \psi \right\vert \mathbf{e}_{\alpha }\mathbf{E}^{(-)}(\mathbf{r})\mathbf{e}_{\alpha }^{\ast }
\mathbf{E}^{(+)}(\mathbf{r})\left\vert \psi
\right\rangle &=&\hbar ^{2}\left\vert b_{o}\right\vert ^{2}\left\vert
\sum\limits_{e,e^{\prime }\neq o}\widetilde{\Sigma }_{\alpha e}(\omega
_{o})R_{ee^{\prime }}(\omega _{o})\Sigma _{e^{\prime }o}(\omega
_{o})\right\vert ^{2}.  \label{39}
\end{eqnarray}

Knowing the transition probability per unit time, we can determine the differential cross section of light scattering by an atomic cloud. We calculate this cross section by assuming that the distance from the source to the cloud is so large that a wave coming from the source to the atomic ensemble is a plane wave with the wave vector $\mathbf{k=}\omega _{o}\mathbf{n/}c$ with good accuracy [see (\ref{27.2})]. This wave is polarized. The direction of its polarization unit vector $\mathbf{e}$ is determined by the orientation of the vector $\mathbf{k}$ and a Zeeman sublevel to which the source atoms was excited. We will find the photon flux density $J$  of this wave in the atomic ensemble region from relation (\ref{37}) by retaining only the first term in it: $\
J=\sum\limits_{\alpha }c\left\vert b_{o}\right\vert ^{2}\hbar \left\vert \widetilde{\Sigma }_{\alpha o}(\omega _{o})\right\vert ^{2}/2\pi \omega _{o}.$ Here, the summation is over any two basis polarizations in a coordinate system determined by vector $\mathbf{k}$.

To obtain the explicit expression for the scattering cross section, we retain in expression (\ref{32}) for the probability $w$ only the contribution caused by the cloud, sum it over a narrow group of finite states of the field, and divide by the flux density $J$ of photons incident on the atomic ensemble from the source. Taking into account that the radiation frequency does not change upon scattering in this case, we obtain the expression
$\mathbf{k}^{\prime }\mathbf{,e}^{\prime } $\begin{equation} \frac{d\sigma }{d\Omega
}=\frac{\omega _{o}^{4}}{\hbar ^{2}c^{4}}\left\vert \sum\limits_{e,e^{\prime }\neq o}\left(
\mathbf{e}^{\prime \ast }\mathbf{d}_{g;e}\right) R_{ee^{\prime }}(\omega _{o})\left(
\mathbf{ed}_{e^{\prime };g}\right) \exp \left( i(\mathbf{kr}_{e^{\prime }}-\mathbf{k}^{\prime
}\mathbf{r}_{e}\right) \right\vert ^{2}.  \label{40}
\end{equation}\qquad \qquad
for the cross section of photon scattering from the mode $\mathbf{k,e}$ , to an arbitrary mode $\mathbf{k}^{\prime }\mathbf{,e}^{\prime } $. 

Note that we obtained the relation for the total cross section earlier \cite{SKKH09} by using the T-matrix formalism and showed, in particular, that the $R_{ee^{\prime }}$ matrix is the projection of the resolvent of the system under study on the states with one excited atom. The relations obtained in this section will be further used to analyze the spectral, angular, and polarization characteristics of light scattering by ultracold atomic clouds.

\section{RESULTS OF CALCULATION}
\setcounter{equation}{0} 
The main quantity necessary for the description of scattering is a resolvent $R_{ee^{\prime }}(\omega )$ because it determines amplitudes (\ref{22.3})-(\ref{26}) of all quantum states involved in the problem and all experimental characteristics of the process under study [see, for example, (\ref{34}),
(\ref{37}) and (\ref{40})). In our paper, this resolvent was found numerically. Depending on convenience, we calculated either the reciprocal matrix (\ref{21.1}) or directly found the numerical solution to system (\ref{19}). The resolvent value considerably depends on the spatial configuration of the ensemble under study. The analysis of random inhomogeneities of a medium and averaging over these inhomogeneities were performed by the Monte Carlo method of statistical tests. The spatial distribution of atoms in the cloud was specified by means of pseudorandom-number generators, and then a particular observable of interest for us was calculated. This observable was calculated repeatedly for different random configurations, and the results were averaged. Note that in most of the papers studying collective effects in dense media, the averaging procedure is performed differently by replacing the real medium by the averaged medium with a continuous density distribution, and then the observables are calculated for this averaged medium. As follows from our analysis \cite{SKKH09}, in this case, some correlation effects affecting observables are described incorrectly.

The expressions obtained in the previous section describe light scattering by ensembles of ultracold atoms under various experimental conditions, in particular, for spatially inhomogeneous clouds of different shapes and arbitrary densities. We are mainly interested in dense clouds with high atomic concentrations in which dependent scattering effects are important. However, we will not restrict ourselves to this case only. Some results obtained in the paper concern optically dense but rarefied atomic ensembles. The analysis of the weak localization regime as a limiting case is important, first, for verifying and substantiating the method proposed and, second, it allows us to refine some results that were earlier obtained in this regime by other methods.

Below, we analyze the two main characteristics of scattering measured in experiments, namely, the angular (spatial) distribution of scattered light and the spectral dependence of the scattering cross section. We restrict ourselves to the results obtained for the quasistatic case.

\subsection{Spatial Distribution of Scattered Radiation: The Effect of Coherent Backscattering}

The scattering of light by an optically dense medium in the presence of localization phenomena is considerably determined by the effects of interference and radiation trapping. Therefore, we verify first of all that the approach proposed in this paper correctly describes these effects.

The basic equations (\ref{19}) do not contain derivatives with respect to spatial variables. In this connection it is interesting to study how these equations describe radiation transfer in an optically dense medium, in particular, how they reproduce the Bouguer-Lambert law. According to this law, the intensity of the coherent component of radiation transmitted through a medium layer is attenuated exponentially. The exponent $b=n\sigma L$ (the optical thickness of the layer) for a homogeneous medium is determined by the concentration $n$ of atoms, the layer thickness $L$, and the scattering cross section $\sigma $ by an atom. For the J = 0 $\leftrightarrow$ J=1 transition in a rarefied medium, we have $\sigma =6\pi \lambdabar^{2}$.

\begin{figure}[th]
\begin{center}
{\large \scalebox{0.55}{\includegraphics*{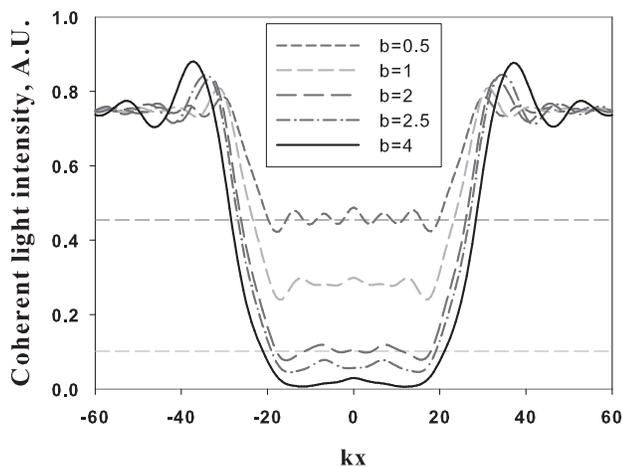}}  }
\end{center}
\caption{Spatial intensity distribution $I_{coh}$ of the coherent component of light scattered by a rarefied cylindrical cloud. Coordinate $x$ is measured in a plane perpendicular to the cylinder axis. The cloud radius is $R=24\lambdabar$ and the concentration is $n\lambdabar^{3}=10^{-3}.$ The different curves correspond to different optical thicknesses $b=n\sigma L$ of the scattering cloud.} \label{f1}
\end{figure}

To analyze the radiation transfer process, we consider the scattering of resonance radiation by a model cylindrical atomic cloud with radius $R$ and length $L$. The wave vector of the incident wave is assumed parallel to the cylinder axis. We will calculate the spatial intensity distribution $I_{coh}$ for the coherent component [which is determined by the square of the amplitude modulus (\ref{34}))] in a plane perpendicular to this axis. The distance $Z_{d}$ from the rear of the cylinder to the observation plane is chosen equal to a few $\lambdabar$ to fit several Fresnel zones within the end for any point of this plane. The results of calculations for a rarefied medium with $n\lambdabar^{3}=10^{-3},\ R=24\lambdabar,$ $Z_{d}=4\lambdabar$, and different $L$ are shown in \ref{f1}.

The obtained intensity distribution corresponds to the Fresnel pattern for diffraction by a semitransparent screen. Outside the shadow region, oscillations are observed typical of diffraction from the screen edge. The intensity attenuation in the geometrical shadow region is caused by the destructive interference of the source radiation and secondary radiation emitted by the atomic ensemble. This interference formally appears in our approach when two terms in (\ref{34}) are taken into account. Note that the diffraction pattern in our approach is obtained directly upon summation of contributions from a great number of microscopic scatterers forming the medium.

\begin{figure}[th]
\begin{center}
{\large \scalebox{0.55}{\includegraphics*{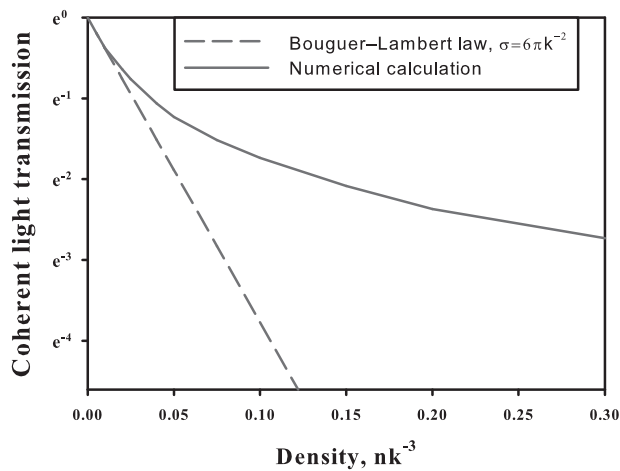}}  }
\end{center}
\caption{Transmission coefficient of the atomic cloud for the coherent component as a function of cloud concentration (solid curve). The deviation from the Bouguer-Lambert law $\sigma =6\pi \lambdabar^{2}$ (dashed straight line) on passing from rarefied to dense clouds is clearly seen.
 } \label{f3}
\end{figure}

The transmission coefficient $T_{coh}$ was calculated by finding the mean light intensity in the geometrical shadow region. These mean intensities for two thicknesses are shown in \ref{f1} by the dashed lines. The dependence of the sample transmission on its length $L$ showed that the Bouguer-Lambert law is reproduced quite accurately in our calculations. In this case, the optical thickness increases linearly with increasing $L$. As for the dependence of optical thickness $b$ on concentration $n$, it increases linearly only up to concentration $n\lambdabar^{3} \leq 10^{-2}$  (\ref{f3}). The deviation from linearity at higher $n$ is caused by collective effects, namely, by the interatomic resonance dipole-dipole interaction. This interaction leads to shifts of atomic levels, so that the scattering cross section  $\sigma $  for resonance radiation decreases. This decrease partially compensates the increase in concentration: the higher the atomic density, the better the compensation. This effect, in particular, can partially explain the difficulty in experimental observation of strong localization in atomic gases.

We now consider light scattering in the case of Fraunhofer diffraction. We analyze the angular distribution of the cloud radiation at a large distance from it. The calculation is performed using expression (\ref{40}) for the differential scattering cross section. We assume that the light frequency coincides with the transition frequency in a free atom, while the probe radiation is circularly polarized. We will analyze, in particular, coherent backscattering (CBS); therefore, to avoid problems with sharp boundaries, the atomic cloud will be assumed inhomogeneous with a Gaussian spatial density distribution.

\begin{figure}[th]
\begin{center}
{\large \scalebox{0.55}{\includegraphics*{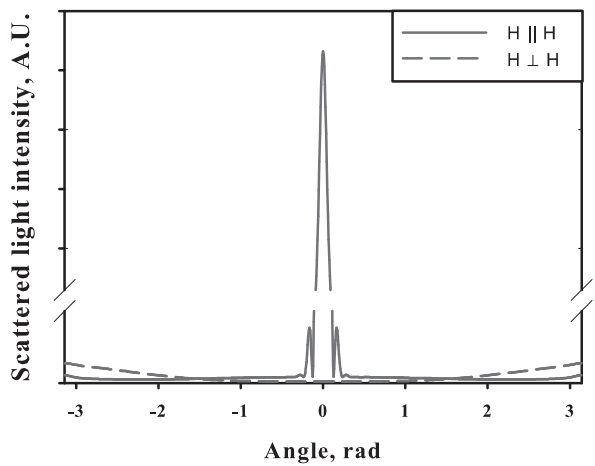}}  }
\end{center}
\caption{Angular distribution of scattered light intensity for two polarization channels. The solid and dashed curves correspond to scattering without a change in the helicity (H $\parallel$ H) and with a change in helicity to the opposite (H $\perp$ H). 
} \label{f4}
\end{figure}

Figure \ref{f4} shows the angular distribution of scattered light for two polarization channels, with light helicity preserved (H $\parallel$ H) or changing to the opposite (H $\perp$ H). The calculations were performed for a cloud with a Gaussian radius $R=12\lambdabar.$. The density at the cloud center was $n\lambdabar^{3}=10^{-2}.$ This corresponds approximately to the intermediate value between dense and rarefied ensembles. When the helicity was preserved, we observed typical Fraunhofer diffraction. The principal diffraction maximum caused by coherent forward scattering dominates. Several higher-order peaks are well distinguished, although because of a smooth decrease in the concentration at the edges of the Gaussian cloud, these peaks are not as distinct as for diffraction from objects with sharp boundaries. Along with diffraction peaks, a comparatively weak background is observed, which is caused by multiple scattering in the cloud. The background intensity weakly depends on angles, except for a narrow region near backward scattering, where a local maximum is observed. This is the CBS cone.

Radiation with orthogonal helicity (the dashed curve in Fig. \ref{f4}) appears due to incoherent multiple scattering (incoherent light scattering by different atoms). The intensity of this radiation in the case under study is low. It increases monotonically for scattering angles above  $\pi /2$. The CBS effect for this polarization channel and the transition under study is known to be a very weak (see, for example, \cite{23})) and is almost unnoticeable at this scale, which is correctly reproduced by our calculations. Thus, our microscopic calculation reproduces the interference effect upon multiple scattering in a disordered medium for both main scattering channels. However, unlike methods used earlier for calculations of this effect and taking into account only a finite number of scattering orders, our approach takes into account all processes of multiple scattering of photons in a medium and also allows us to determine the cone shape at arbitrary atomic concentrations.

\begin{figure}[th]
\begin{center}
{\large \scalebox{0.55}{\includegraphics*{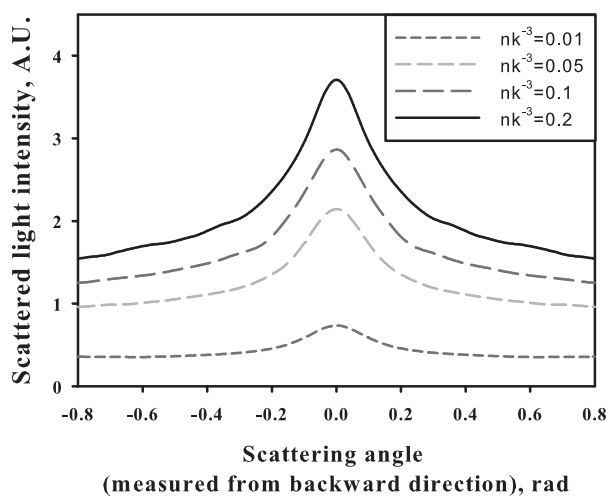}}  }
\end{center}
\caption{Cone shape for CBS by a spherically symmetric Gaussian cloud with a radius of $12\lambdabar$ as a function of the atomic concentration 
 } \label{f5}
\end{figure}

The CBS properties are shown in more detail in Fig. \ref{f5}. This figure demonstrates the concentration dependence of the CBS cone for the H $\parallel$ H channel. The visible shape of the cone changes comparatively weakly in this range of $n$. Figure 4 demonstrates the qualitative change in the CBS effect for dense clouds. It is known that all weak-localization theories proposed so far predict the maximum value of the amplification factor (the ratio of the cone maximum to the value of the quasi-isotropic pedestal) equal to two. This is explained by the fact that the cone amplitude is caused by the interference of the waves scattered by the same chain of atoms but propagating along this chain in opposite directions. These waves in rarefied media interfere only in pairs. We see that our calculation for rarefied media (the lower curve in Fig. \ref{f5}) quite accurately gives an amplification factor of 2. The amplification of backscattering in denser media exceeds 2. In our opinion, this is explained by two main reasons. First, in dense media, along with the conventional mechanism resulting in weak localization, the interference of beams scattered by different but spatially close chains makes a certain contribution. The second reason is the decrease in the fraction of light scattered outside the cone at comparatively large angles, i.e., the decrease in the background intensity with respect to which the amplification effect is measured (see details in \cite{FABSAMM}).

\subsection{Scattering Spectrum}

We consider the spectral dependences of the differential and total scattering cross sections as another example of the application of the theory developed above. The total cross section can be obtained by integrating the differential cross section over the entire spherical angle or, according to the optical theorem, by calculating the amplitude of light scattered at the zero angle. In this paper, we used both these methods, the optical theorem being fulfilled independently of the cloud density.

\begin{figure}[th]
\begin{center}
{\large \scalebox{0.55}{\includegraphics*{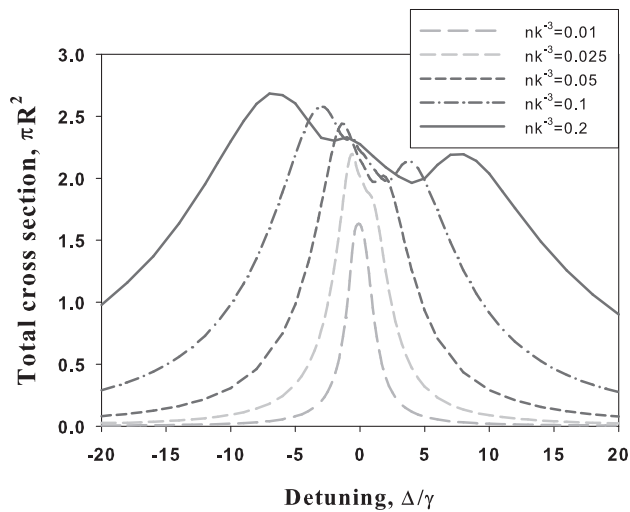}}  }
\end{center}
\caption{Spectra of the total cross section of light scattered by a spherical cloud with radius $R=15\lambdabar$ for different atomic concentrations. 
 } \label{f6}
\end{figure}

Figure \ref{f6} shows the total scattering cross section spectra for a spherically homogeneous, on average, cloud with radius $R=15\lambdabar$ at different densities. At low concentrations, when the interatomic interaction is insignificant, we see a monotonic spectral dependence, which is typical of scattering by an isolated atom. The spectrum is somewhat broadened compared to the monatomic scattering spectrum because this medium is not optically thin even at these concen trations. As the cloud density is increased, the shape of the spectrum begins to be distorted. At the concentration $n\lambdabar^{3}=0.2$, the spectral dependence is substantially nonmonotonic, and several distinct local maxima appear.

This behavior of the cross section can be explained either in the context of macroscopic electrodynamics or by using the microscopic approach. Macroscopic electrodynamics describes scattering as the result of multibeam interference (or diffraction) from a sphere with the specified permittivity. Such scattering in the case of spatially homogeneous spheres with arbitrary permittivities was considered in detail in the Debye- Mie theory (see, for example, \cite{BV}). This theory showed, in particular, that the maximum scattering cross section for transparent spheres with the refractive index $\mathfrak{n}$ was observed for $2\widetilde{R}k(\mathfrak{n}-1)\simeq 4. $. In this case, the scattering cross section was approximately four times greater than the sphere cross section area
$\pi R^{2}$. As the radius is decreased compared to $R$, the ratio of the scattering cross section $\sigma $ to $\pi R^{2}$ tends to zero. As $R$ is increased, the ration $\sigma $/$\pi R^{2} $  oscillates and tends to 2 for $R\rightarrow \infty .$ The scattering cross section behaves similarly when the sphere material has a lower optical density than the environment. However, there exists some quantitative difference. The maximum scattering cross section is lower and is observed for smaller values of $R$. If the medium is not transparent, i.e., the refractive index has a noticeable imaginary part (the absorption coefficient), the diffraction type changes. The oscillations mentioned above are considerably smoothed, although the maximum scattering cross section can exceed $2\pi R^{2}.$

This behavior of the scattering cross section agrees qualitatively with results shown in Fig. \ref{f6}. In the case of an atomic cloud, the permittivity is complex. We can introduce the real refractive index and absorption coefficient (although note that in the case under study, a wave is not really absorbed by an atomic ensemble because the escape of photons from the incident wave-mode is accompanied by the appearance of photons in other modes). The microscopic approach developed in our paper gives accurate values for the permittivity of a homogeneous atomic medium with the specified density. However, the discussion of this issue is beyond the scope of this paper. Note that the permittivity value is determined by studying the spatial distribution of the atomic polarization taking into account that for detunings $\Delta $ exceeding the atomic resonance width, the refractive index decreases with increasing $\Delta $ considerably slower than the absorption coefficient. For the specified radius of the cloud $Rk=15$, the condition $2Rk(\mathfrak{n}-1)\simeq
\mathfrak{n}-1\simeq 0.1333.$

Our calculation of the permittivity shows that the position of the left maximum qualitatively agrees with the predictions of the Mie theory. The intensity of this maximum is smaller than the maximum possible value for transparent media, which is related to a small but finite absorption coefficient in this frequency region. As the atomic concentration in the cloud is increased, the maximum shifts to the region of greater detuning magnitudes and its amplitude increases. This is explained by the fact that a comparatively small value of $\mathfrak{n}$ in a denser medium is achieved for a large magnitude of detunings $\Delta $. The relative influence of absorption for large detunings also decreases.

The maximum observed for positive detunings also can be explained by the Debye-Mie theory. The refractive index of the atomic medium in this frequency region is smaller than unity, and therefore the right maximum intensity is lower and it is located at smaller magnitudes of $\Delta $ than the left one.

The absorption coefficient for small detunings is large compared to $\mathfrak{n}-1$ (the difference of the real refractive index from unity), and therefore there exists another peak related to the absorption coefficient maximum. This peak is most distinctly observed at large concentrations when the spectral positions of all three cross section maxima are well distinguished.

\begin{figure}[th]
\begin{center}
{\large \scalebox{0.55}{\includegraphics*{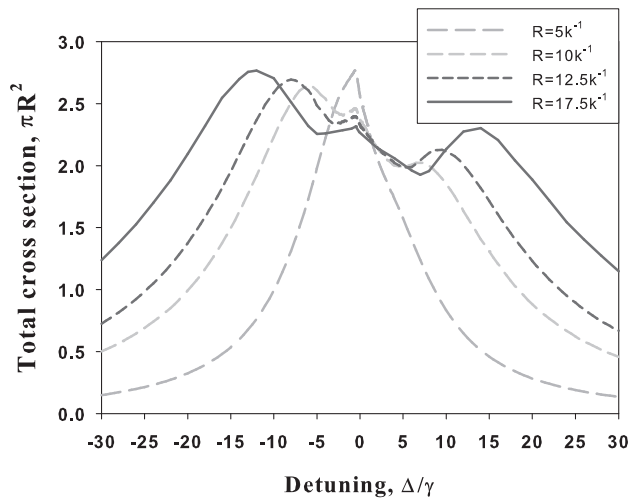}}  }
\end{center}
\caption{Spectra of the total cross section of light scattered by spherical clouds of different sizes. The atomic concentration is $n\lambdabar^{3}=0.1$.
 } \label{f7}
\end{figure}

Figure \ref{f6} shows the concentration dependence of the shape of the total cross section spectrum for a fixed cloud size. Figure \ref{f7} illustrates the dependence of the cross section on cloud size for the specified concentration. The calculation was performed for $n\lambdabar^{3}=0.3$ when the interatomic interaction is considerable. We can see that, as the cloud size increases, the maxima caused by refraction shift to the region of large detunings, according to the condition $\mathfrak{n}-1\simeq 2/Rk.$ The position of the central maximum at the specified concentration does not change.

The frequency dependence of the total cross section can be explained by using the microscopic approach. To do this, it is necessary to consider the distributions of collective atomic states over energies and lifetimes, more exactly, the distribution of the poles of the resolvent $R_{ee^{\prime }}(\omega )$ over the values of their real and imaginary parts 

\footnote{This refinement is required because the presence of resonance dipole-dipole interaction in a system of atoms prevents the introduction of collective states decaying independently (the matrix in the left-hand side of equation (\ref{19}) is not normal, and its real and imaginary parts cannot be simultaneously reduced to the diagonal form). Our analysis has shown that the maximum of the spectral dependence corresponded to the maximum density of such states taking into account their finite width. Note here that the study of poles gives the position of maxima more accurately than the Debye-Mie method. This is explained by the fact that the microscopic approach correctly takes into account boundary effects. With the parameters under study, atoms in boundary regions are located in different conditions than atoms in central parts because they interact with a smaller number of neighbors, whereas the permittivity in the Debye-Mie theory is assumed the same over the entire volume of the sphere}.

\begin{figure}[th]
\begin{center}
{\large \scalebox{0.55}{\includegraphics*{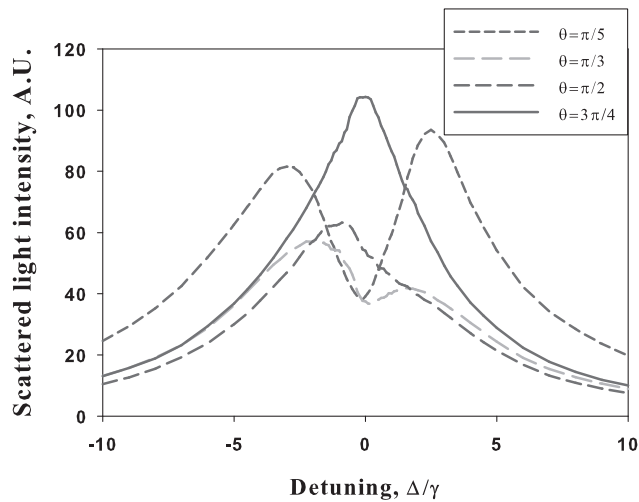}}  }
\end{center}
\caption{Intensity of light scattered in different directions as a function of the incident radiation frequency. 
 } \label{f8}
\end{figure}

In light scattering experiments, the differential cross section is most often studied rather than the total cross section. As a rule, the intensity of light scattered through a specified angle is measured \cite{BWHSK} and its dependence on the probe light frequency is determined. Figure \ref{f8} shows this dependence calculated by us for a spherical cloud with radius $Rk=15$ and concentration $nk^{-3}=0.1.$ On the ordinate, the total light intensity (without selection over polarizations) is plotted. Different curves correspond to scattering through different angles ?. Note that the behavior of these curves is substantially different. For some scattering angles, the spectral dependence is monotonic, whereas for others, the nonmonotonic behavior is observed. A similar difference will be manifested in the fluorescence of the cloud excited by a light pulse. In the case of a short pulse whose spectrum overlaps the entire excitation spectrum of the atomic ensemble, fluorescence in some directions will decay monotonically, while in other directions, beats will be observed.

\begin{figure}[th]
\begin{center}
{\large \scalebox{0.55}{\includegraphics*{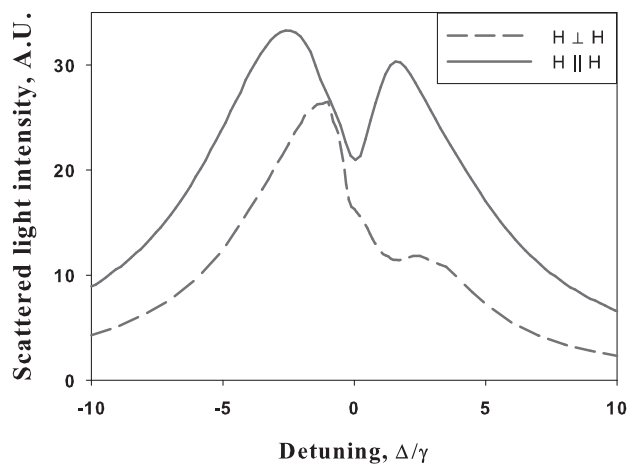}}  }
\end{center}
\caption{Spectral dependences of the intensity of light scattered through angle $\pi /3$ for two polarization channels: solid curve and dashed curves are scattering without a change in the helicity H $\parallel$ H  and with a change in the helicity to the opposite H $\perp$ H.
 } \label{f9}
\end{figure}

The spectral behavior is different not only for scattering in different directions but also for different polarization components scattered in the same direction. This is illustrated in Fig.  \ref{f9}, which shows the intensities of two orthogonal polarization components as functions of the incident light frequency. The calculation was performed for the same conditions as in Fig. \ref{f8}. The scattering angle is $\pi /3$. The polarization of incident light, as in all other calculations, was circular. The results demonstrate a considerable dispersion of polarization: the polarization of light scattered in the given direction can considerably depend on the incident light frequency.

\section{CONCLUSIONS}

We have developed a consistent quantum-mechanical theory of cooperative scattering of weak electromagnetic radiation by an optically dense macroscopic atomic ensemble. The microscopic approach proposed in the paper can be applied both to rarefied clouds, when the average number of atoms in the volume is small, and to the clouds in which this parameter exceeds unity. The differential and total dependent scattering cross sections have been calculated for quasi stationary radiation using this approach. The angular, spectral, and polarization properties of scattered light were determined. The dependence of these characteristics on cloud size and atomic concentration was studied. By summing the contributions from many atomic scatterers, we obtained, along with the known Fresnel and Fraunhofer diffraction phenomena, the contribution from multipole incoherent scattering and calculated the interference effect of coherent backscattering. The parameters of the backscattering cone were determined. Note that, unlike approximate methods used earlier, in which backscattering was calculated taking into account a finite number of scattering events, our approach allows us to calculate these parameters exactly. This effect was calculated for the first time for dense clouds where the mean free path is comparable with the light wavelength. It was found that the complex nature of interference effects, when aside from the traditional mechanism resulting in a weak localization, it is necessary to consider the interference of beams scattered by closely located chains, leads to an amplification factor exceeding 2.

The dependence of the scattering cross section on the incident light frequency has demonstrated a strong modification of the spectra with increasing atomic concentration. It was shown that these spectra for dense media have a prominent nonmonotonic nature and exhibit several maxima. It was also found that collective effects lead to a considerable difference in the spectral dependences of scattering in different directions. The investigation of the polarization properties of scattered light revealed a considerable dispersion of polarization, which strongly depends on the incident light frequency.

Our study concerns the case when the motion of atoms can be neglected. It is obvious, however, that this motion can play an important role in some experiments. Therefore, we believe that it is interesting to consider further the influence of atomic motion on collective effects in dense atomic systems and the interaction of light with such systems. In addition, it is important for practical purposes to generalize the microscopic approach to the case when an additional control field is present. In our opinion, the use of high density atomic systems to obtain electromagnetically induced transparency and to "stop light" can be quite efficient for solving problems in quantum informatics.

\begin{acknowledgments}
This work supported by the Russian Foundation for Basic Research (project nos. 08-02-91355  and 10-02-00103) and the United States National Science Foundation (grant no. NSF-PHY-0654226).
\end{acknowledgments}

\section{Appendix}
\setcounter{equation}{0}

Infinite sums $\sum\limits_{g}V_{eg}V_{ge^{\prime }}\varsigma
\left( \hbar \omega -E_{g}\right) $ and $\sum\limits_{ee}V_{e;ee}V_{ee;e^{\prime }}\varsigma \left( \hbar
\omega -E_{ee}\right) $ appearing in Eqs. (\ref{16}) can be calculated for an infinitely broadband field thermostat assuming that the quantization volume is infinite. We first consider the first sum, assuming that in state $|e^{\prime }\rangle $ the atom $a$ is excited, while in  $|e\rangle $
 atom $b$ is excited. In dipole approximation (\ref{3}), we have
\begin{gather}
\sum\limits_{g}V_{eg}V_{ge^{\prime }}\varsigma \left( \hbar \omega -E_{g}\right)=-
\sum\limits_{\mu ,\nu }\mathbf{d}_{e_{b};g_{b}}^{\mu }\mathbf{d}_{g_{a};e_{a}}^{\nu }
\underset{\sigma \rightarrow 0}{\lim }\int_{0}^{\infty }\left( \delta _{\mu \nu }\nabla
^{2}-\frac{\partial ^{2}}{\partial \mathbf{r}_{\mu }\partial \mathbf{r}_{\nu }}\right)
\frac{\,\,cdk}{\pi r}\frac{\sin (kr)}{\omega -kc+i\sigma }. \label{A.4}
\end{gather}
Here, $\mathbf{r}=\mathbf{r}_{b}\mathbf{-r}_{a}$. Expression (\ref{A.4}) was derived by calculating the sum over polarizations using completeness relation (\ref{5}), replacing the sum over $\mathbf{k}$ by an integral, and using one of the representations of the singular function $\varsigma(x)$.

Formally, some of the integrals in (\ref{A.4}) diverge. However, this divergence is related to the behavior of integrands for large $k$, which is caused by the use of the dipole approximation. By introducing, as usual, a slowly decaying exponential factor$\exp (-\varepsilon k)$ with a small exponent $\varepsilon
>0,$, we can provide the divergence of integrals. By calculating integrals and tending the parameter $\varepsilon $ to zero in finite expressions, we obtain
\begin{equation} \sum\limits_{g\neq o}V_{eg}V_{ge^{\prime
}}\varsigma \left( \hbar \omega -E_{g}\right) =\sum\limits_{\mu ,\nu
}\mathbf{d}_{e_{b};g_{b}}^{\mu }\mathbf{d}_{g_{a};e_{a}}^{\nu
}\left[ \delta _{\mu \nu }F_{1}(\frac{\omega
r}{c})+\dfrac{\mathbf{r}_{\mu }\mathbf{r}_{\nu
}}{r^{2}}F_{2}(\frac{\omega r}{c})\right] ;  \label{A.8}
\end{equation}Where \begin{gather}
\pi r^{3}F_{1}(\frac{\omega r}{c})=-\left( \sin \frac{\omega
r}{c}\left( -i\pi +Ci\left( \frac{\omega r}{c}\right) \right) -\cos
\frac{\omega r}{c} \left( \frac{\pi }{2}+Si\left( \frac{\omega
r}{c}\right) \right)
\right) +  \notag \\
+\frac{\omega r}{c}\left( \cos \frac{\omega r}{c}\left( -i\pi +Ci
\left( \frac{\omega r}{c}\right) \right) +\sin \frac{\omega
r}{c}\left( \frac{\pi }{2}+
Si\left( \frac{\omega r}{c}\right) \right) \right) - \notag \\
-\frac{\omega r}{c}+\frac{\omega ^{2}r^{2}}{c^{2}}\left( \sin \frac{\omega r}{c}\left( -i\pi +Ci\left( \frac{\omega r}{c}\right) \right) -\cos \frac{\omega r}{c}\left( \frac{\pi }{2}+Si\left( \frac{\omega r}{c}\right) \right) \right) ;  \label{A.9} \\
\pi r^{3}F_{2}(\frac{\omega r}{c})=3\left( \sin \frac{\omega
r}{c}\left( -i\pi +Ci\left( \frac{\omega r}{c}\right) \right) -\cos
\frac{\omega r}{c}\left( \frac{\pi }{2}+Si\left( \frac{\omega
r}{c}\right) \right)
\right) -  \notag \\
-3\frac{\omega r}{c}\left( \cos \frac{\omega r}{c}\left( -i\pi +Ci\left( \frac{\omega r}{c}\right) \right) +\sin \frac{\omega r}{c}\left( \frac{\pi }{2}+Si\left( \frac{\omega r}{c}\right) \right) \right) + \notag \\
+\frac{\omega r}{c}-\frac{\omega ^{2}r^{2}}{c^{2}}\left( \sin
\frac{\omega r}{c}\left( -i\pi +Ci\left( \frac{\omega r}{c}\right)
\right) -\cos \frac{\omega r}{c}\left( \frac{\pi }{2}+Si\left(
\frac{\omega r}{c}\right) \right) \right) . \label{A.10}
\end{gather}
Here, $Ci(x)$ and $Si(x)$ are integral cosine and sine, respectively.

Similarly, we can calculate the two-atom contribution to the second sum caused by nonresonance intermediate states
\begin{equation} \sum\limits_{ee}V_{e;ee}V_{ee;e^{\prime
}}\varsigma \left( \hbar \omega -E_{ee}\right) =\sum\limits_{\mu
,\nu }\mathbf{d}_{e_{b};g_{b}}^{\mu }\mathbf{d}_{g_{a};e_{a}}^{\nu
}\left[ \delta _{\mu \nu }F_{1}\left( \frac{\left( \omega -2\omega
_{a}\right) r}{c}\right) +\dfrac{\mathbf{r}_{\mu }\mathbf{r}_{\nu
}}{r^{2}}F_{2}\left( \frac{\left( \omega -2\omega _{a}\right)
r}{c}\right) \right] ;  \label{A.101}
\end{equation}
Here $\omega _{a}$ is the transition frequency of atoms forming the cloud. It is easy to see that this contribution for small $r$ is comparable with resonance term (\ref{A.8}).

Expressions (\ref{A.8}) and (\ref{A.101}) can be considerably simplified in the so called polar approximation, when their value for frequency $\omega $ is replaced by their value for frequency$\omega _{a}$ of the atomic resonance. This approximation was studied in detail in \cite{MK74}, where it was shown that it can be applied in systems where retardation effects are insignificant. The value of $R$ for atomic clouds in quasistatic traps, which are studied in this paper, does not exceed hundreds of micrometers. The shortest evolution times of the atomic system to be compared with the delay times are the lifetimes of short-lived superradiance states. For systems with size $R$, these times are on the order of $\tau (nk^{-2}R)^{-1}$  where $\tau $ is the natural excited state lifetime of a free atom. Under such conditions and for typical parameters of alkali metal atoms, retardation effects can be neglected even for clouds with density $nk^{-3}\sim 1.$ In this case, using the polar approximation, we obtain
\begin{eqnarray} F_{1}\left( \frac{\left( \omega
-2\omega _{a}\right) r}{c}\right) +F_{1}\left( \frac{\omega
_{a}r}{c}\right) &=&\frac{1}{r^{3}}\left( 1-i\frac{\omega
_{a}r}{c}-\left( \frac{\omega _{a}r}{c}\right) ^{2}\right) \exp
\left( i\frac{\omega _{a}r}{c}\right) ;  \label{A.11} \\
F_{2}\left( \frac{\left( \omega -2\omega _{a}\right) r}{c}\right)
+F_{2}\left( \frac{\omega _{a}r}{c}\right) &=&-\frac{1}{r^{3}}\left( 3-3i\frac{\omega _{a}r}{c}-\left( \frac{\omega _{a}r}{c}\right) ^{2}\right) \exp
\left( i\frac{\omega _{a}r}{c}\right)  \label{A.12}
\end{eqnarray}
\begin{eqnarray*}
\Sigma _{ee^{\prime }}(\omega ) &=&=\frac{\mathbf{d}_{e_{b};g_{b}}^{\mu }\mathbf{d}_{g_{a};e_{a}}^{\nu }}{\hbar r^{3}}\left[ \delta _{\mu \nu }\left(
1-i\frac{\omega _{a}r}{c}-\left( \frac{\omega _{a}r}{c}\right) ^{2}\right)
\exp \left( i\frac{\omega _{a}r}{c}\right) \right. + \\
&&\left. -\dfrac{\mathbf{r}_{\mu }\mathbf{r}_{\nu }}{r^{2}}\left( 3-3i\frac{\omega _{a}r}{c}-\left( \frac{\omega _{a}r}{c}\right) ^{2}\right) \exp
\left( i\frac{\omega _{a}r}{c}\right) \right] .
\end{eqnarray*}
Note that the latter expression can be also rewritten in terms of spherical Hankel functions and spherical harmonics \cite{SKKH09,Berman}.

The monatomic contribution, when $e^{\prime }=e,$ has the form
\begin{equation}
\sum\limits_{g}V_{eg}V_{ge}\varsigma \left( \hbar \omega -E_{g}\right) =-\frac{i\hbar \gamma }{2}  \label{A.13}
\end{equation}
where $\gamma $  is the spontaneous decay rate, which is the same for all the sublevels of the excited state. 

The sum $\sum\limits_{ee}V_{e;ee}V_{ee;e}\varsigma \left( \hbar \omega -E_{ee}\right) $ gives only the Lamb shift of
the ground state  because it describes the spontaneous excitation of an atom located close to the excited atom, which is accompanied by spontaneous decay with absorption of a virtual photon. We neglect the Lamb shifts of the ground and excited states, assuming that they are included in the transition frequency $\omega _{a}.$

\bigskip

\end{document}